\documentclass[reprint, amsmath,amssymb,pra]{revtex4-2}

\usepackage{graphicx}% Include figure files
\usepackage{dcolumn}% Align table columns on decimal point
\usepackage{bm}% bold math
\usepackage[unicode=true,pdfusetitle,bookmarks=false,breaklinks=false,pdfborder={0 0 0},backref=false,colorlinks=false]{hyperref}

\usepackage[textwidth=1cm]{todonotes}

\begin{document}

\title{Picosecond laser pulses for quantum dot--microcavity based single photon generation by cascaded electro-optic modulation of a narrow-linewidth laser}

\author{Mio Poortvliet}
\email{poortvliet@physics.leidenuniv.nl}
\author{Petr Steindl}
\author{Ilse Kuijf}
\author{Harry Visser}
\author{Arno van Amersfoort}
\author{Wolfgang L\"offler}

\affiliation{%
Leiden Institute of Physics, Leiden University, 2333 CA Leiden, The Netherlands
}%

\date{\today}

\begin{abstract}
    Recent developments in integrated optics have made it possible to fabricate high-bandwidth electro-optic modulators (EOMs). Here we show cascaded operation of two of such modulators driven by custom-built electronics delivering on-demand picosecond pulses and application to a quantum dot cavity-QED single photon source. We implement an EOM-based correlation technique and demonstrate light pulses as short as $24\pm2$ ps with a single EOM. The EOMs can be synchronized and operated in series, we then produce optical pulses down to $17$ ps. To optimize the pulse contrast, we analyze for two different EOM devices the transmission as a function of EOM bias, wavelength and temperature, and we show that by temperature tuning and stabilization, a pulse contrast above 25 dB can be obtained at the targeted wavelength. With this high contrast we demonstrate resonant excitation of an InGaAs quantum dot - microcavity based single photon source, demonstrating a crucial technology for scalability and synchronization of large scale photonic quantum applications.
\end{abstract}

\maketitle

\section{Introduction}
Fast, flexible and indistinguishable single photon sources are essential for scalable quantum communication and computing \cite{wehner_quantum_2018}, 
and semiconductor quantum dots (QDs) in optical microcavities are an excellent candidate \cite{somaschi_near-optimal_2016}. 
For advanced real-world applications single photons should be generated with low timing jitter and on-demand, meaning triggered by a signal from the user. Self-assembled InGaAs QDs have the advantage of sub-nanosecond lifetimes, in particular if Purcell enhanced in an optical cavity \cite{somaschi_near-optimal_2016}, allowing the generation of single photons at GHz rates. 

For resonant excitation of a QD-based single photon source, high-contrast bandwidth-limited optical excitation pulses much shorter than the QD lifetime are needed to avoid re-excitation \cite{hanschke_quantum_2018, javadi_cavity-enhanced_2023, huber_optimal_2015, bozzio_enhancing_2022, ollivier_hong-ou-mandel_2021}. The standard technique to produce such pulses, usually around $10$ ps in length, is using pulsed lasers such as Ti:Sa oscillators in combination with a pulse shaper.
The resulting photons reach high photon purity and indistinguishability, however, this method lacks flexibility: while the pulses are made deterministically, they are not on-demand and often at a fixed rate of $\sim$ 80 MHz \cite{dada_indistinguishable_2016, meyer_leveraging_2023}.
Additionally, pulse shaping optics requires mechanical alignment and adjustment to change the pulse length, decreasing flexibility and robustness, in particular important for more complex pulse sequences, such as used for spin-photon entanglement \cite{costeHighrateEntanglementSemiconductor2023} or novel excitation schemes \cite{karliSUPERSchemeAction2022}.

To excite slower systems (e.g. color centers or trapped atoms), it is common to use an electro-optic modulator (EOM) to produce pulses from a continuous-wave laser \cite{hermansEntanglingRemoteQubits2023, hofer_bias_2017}.
Recent developments in integrated photonics have enabled high-bandwidth EOMs based on waveguide Mach-Zehnder modulators. 
Such waveguide-based systems allow very narrowly spaced electrodes that require a signal of only a few volts for full $\pi$ modulation, i.e., to switch the output from low to high transmission \cite{wooten_review_2000, valdez_100_2023}. 

Since then, EOM-produced laser pulses have been explored for QD excitation \cite{dada_indistinguishable_2016, fischer_pulsed_2018, hanschke_quantum_2018} with pulse lengths down to $\sim 100$ ps. Dada \textit{et al.} use $\geq100$ ps long pulses for a bare QD, and state that ``much faster pulses ($<$30 ps) can be achieved in the future''  \cite{dada_indistinguishable_2016}. This is realized to some degree in \cite{meyer_leveraging_2023}, however such pulses have not been reported to excite a QD microcavity system, as far as we know. 
We note that EOMs in combination with amplifiers and pulse compression recently enabled generation of even sub-ps pulses, but the added complexity is significant and a high pulse contrast is challenging \cite{renard_agile_2022}.

In this paper we first present our custom-built electronic pulse compressor driven by a field-programmable gate array (FPGA) capable of creating laser pulses down to $24\pm2$ ps length with an EOM from continuous-wave laser light (we report pulse width by the full-width at half maximum, FWHM). Using the Hanburry Brown-Twiss (HBT) technique with single-photon detectors, we analyze the pulse contrast of two different EOM devices as a function of temperature and wavelength. We then introduce a cross-correlation method to characterize ultrashort pulses using a second EOM and only a slow photo diode. After this characterization, we use the EOM-produced laser pulses to resonantly excite an InGaAs QD in an optical microcavity \cite{steindl_resonant_2023, steindl_cross-polarization-extinction_2023}. We show that by cascading two EOMs we can improve the pulse contrast \cite{hofer_bias_2017} and decrease the pulse length \cite{bhattacharya_picosecond_2014}, and demonstrate the effect of pulse contrast on the single-photon purity. Finally, we use our slow-detector cross-correlation method to measure the time-resolved resonance fluorescence of the quantum dot.

\section{EOM and pulse compressor}
\label{sec:eomandcompressor}
We use two LiNbO3 integrated-optics EOMs (both EOMs Exail/iXblue NIR-MPX950-LN-20, one purchased in 2020 and one in 2022), based on a balanced Mach-Zehnder interferometer (MZI). The relative phase between its arms can be modulated using the electro-optic effect in lithium niobate via both a fast (RF) and a slow (DC, bias control) electrode. We use the bias input to operate the EOM in a transmission minimum $T_0$. The RF port has a bandwidth of $>20$ GHz and is used to create the pulses with time-dependent transmission $T(t)=T_0+T_\text{RF}(t).$ An ideal balanced MZI would produce perfect destructive interference ($T_0=0$) but in a real device, slight imbalance of the directional couplers (splitters) results in a finite pulse contrast, specified to $>20$ dB ($T_0<10^{-2}$).

\begin{figure}
    \centering
    \includegraphics[width=\linewidth]{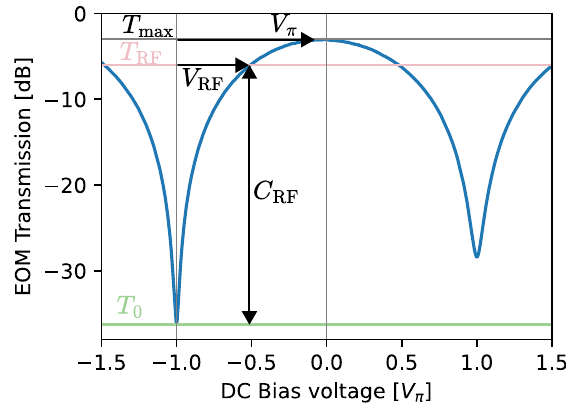} 
    \caption{Transmission of EOM A as function of the DC bias voltage. Notably, one minimum $T_0$ is lower than the other. 
    The RF pulse with voltage $V_\mathrm{RF}$ shifts the transmission away from the minimum, but even when $V_\mathrm{RF}$ does not reach $V_\pi$ exactly, the pulse contrast $C_\text{RF}$ can still be high.}
    \label{fig:EOM-transmission}
\end{figure}

Figure \ref{fig:EOM-transmission} shows the measured EOM transmission as the bias voltage is changed. The voltage required to go from a transmission minimum to a maximum is $V_\pi,$ which is of the order of a few volts, though different for the RF and DC ports. We define the DC contrast as the ratio of the peak $T_\mathrm{max}$ and minimum $T_0$ transmission of the EOM by adjusting the DC bias voltage $V_\pi$. The maximum pulse contrast $T_\mathrm{RF}/T_0$ can be lower due to a reduced $V_\mathrm{RF}$ limited by the electronics. Crucially, the transmission is very sensitive to the bias voltage, and we emphasize that for optimal contrast an optimal bias voltage is more important than reaching $\pi$ modulation by $V_\mathrm{RF}$. 
 
\begin{figure}
    \centering
    \includegraphics[width=\linewidth]{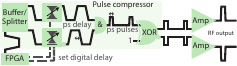}
    \caption{Sketch of the custom RF electronics. An FPGA board delivers a digital pulse to a buffer/splitter, whose two outputs are fed into separate electronic delay lines. A fast AND-gate creates the ps pulses from the (inverted) output of the delay chips, and an ultra-fast XOR gate is used to sharpen the pulse edges. The fast XOR gate has an inverted and an non-inverted output which each can drive an EOM after amplification. }
    \label{fig:EOM-series}
\end{figure}

\begin{figure}
    \centering
    \includegraphics[width=\linewidth]{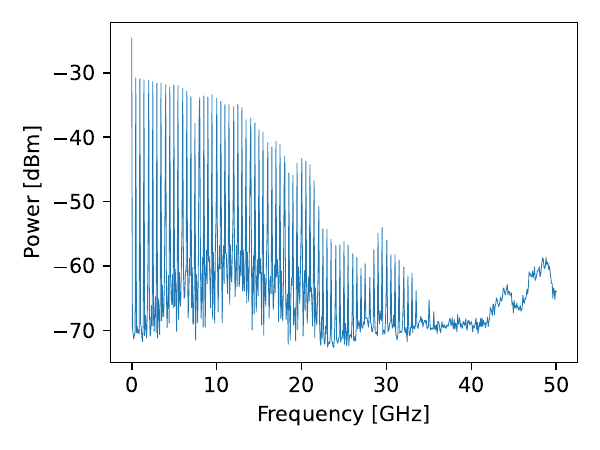}
    \caption{Spectrum analyzer measurement of the output of the pulse compressor board without amplifiers using a pulse distance of $2$ ns and a pulse compressor delay of -75 ps. The resolution bandwidth was set to $30$ kHz and the signal attenuation to $10$ dB to operate in the linear regime of the spectrum analyzer. Due to the periodicity in the pulses there are peaks, which make an envelope that describes the pulse. A fit to this envelope gives a pulse length of $32.7\pm0.8$ ps.}
    \label{fig:spectrum-analyzer}
\end{figure}

The ultra-short electronic pulses for the EOM are created by logical conjunction of two slow pulses in an AND gate as schematically shown in Fig. \ref{fig:EOM-series}. First, we use an FPGA to generate a trigger signal in the form of a pulse pattern with $1$ ns timing resolution, which is sent to the custom pulse compressor board. Here, the trigger signal is copied by a buffer/splitter and sent to two programmable delay lines (Microchip SY89297U) with $5$ ps resolution. The non-inverted output of one delay line and the inverted output of the other are combined in a 14 Gbps AND-gate (Analog Devices HMC726LC3C), producing pulses with a length determined by the relative delay setting of the delay lines. These pulses are fed into a 28 Gbps XOR-gate (Analog Devices HMC851LC3C) with one output permanently pulled high to sharpen the signal edge. Because the fast logic chips work with low signal voltages, an amplifier (Analog Devices HMC994APM5E) is required to achieve $V_{\pi}$. The amplifiers are connected with high-bandwidth 2.92 mm (k) coupling connectors (no cables) followed by a bias tee (not depicted in Fig. \ref{fig:EOM-series}), and the EOM is directly connected to its output. The inverting output of the XOR gate can be used to synchronously drive a second EOM using the same pulse compressor board. The expected pulse shape is rectangular, but in particular for very short pulse lengths ($\lesssim100$ ps) the bandwidth limitations of the components will result in Gaussian profiles. 

The electronic pulses can be verified by a spectrum analyzer measurement as shown in Fig. \ref{fig:spectrum-analyzer} using a $-75$ ps relative delay of the pulse compressor board, which is the delay that results in the shortest measurable pulses. In the frequency domain many harmonics of the $500$ MHz pulse frequency are seen above the noise floor. A Gaussian fit to the envelope results in a pulse length of $32.7\pm0.8$ ps. Since the transmission of the EOM is sinusoidal as a function of voltage (Fig. \ref{fig:EOM-transmission}), it is approximately quadratic around the transmission minimum, $T_{RF}(V_{RF}) \sim V_{RF}^2$, and the expected optical pulse length based on the spectrum analyzer result is $2^{-1/2}\cdot \left( 32.7\pm0.8 \right) \mathrm{ps} = 23.1\pm0.6$ ps.

\begin{figure}
    \centering
    \includegraphics[width=\linewidth]{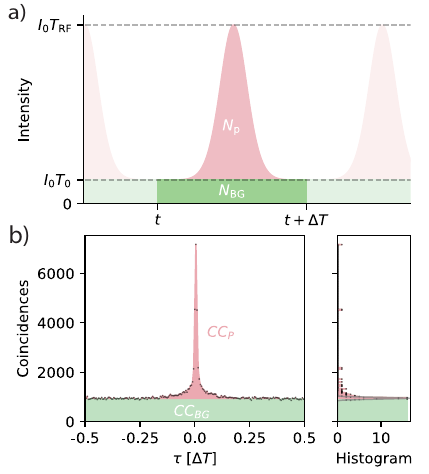} 
    \caption{Time trace of the optical intensity in a pulse (a, theory), and corresponding photon coincidences (b, measured). Periodic pulses with period $\Delta T$ contain each $N_P$ photons on top of a background of $N_{BG}$ photons per period. In a HBT measurement (b), this translates to $CC_P$ coincidence detection events in the pulse on top of a $CC_{BG}$ background coincidences. Due to detector jitter the peak broadens and it is not possible to directly determine the (peak) pulse contrast, but a Gaussian fit to the histogram (b) works well as explained in the text.}
    \label{fig:pulses}
\end{figure}
For quick analysis of the pulse contrast we first introduce some parameters in Fig. \ref{fig:pulses} (a): The number of photons $N_P$ per pulse, peak intensity $I_0 T_{RF}$, on top of a background (leakage) intensity $I_0 T_0$ which contains $N_{BG}$ photons per period $\Delta T$. 
From this we can calculate the ratio of photons in the pulse $N_p$ to the background photons $N_{BG}$. Often, more important is the pulse contrast, $T_{RF}/T_0$, for which we use HBT measurements with 45 ps jitter single photon avalanche photodiodes (SPAPDs) shown in Fig. \ref{fig:pulses} (b).  To determine the pulse contrast, we average over many periods following the approach detailed in Appendix \ref{appx:hbt_analysis}. In Fig. \ref{fig:pulses} (b) we show an exemplary measurement, resulting in a ratio of photons $N_{p}/N_{BG}=0.63\pm0.04.$
The pulse contrast can be estimated to $\sim21$ dB by assuming a square pulse and dividing by the duty cycle of $0.005,$ which was measured using the cross-correlation technique explained in Sec. \ref{sec:xcor}. 

%%%%%%%%%%%%%%%%%%%%%%%%%%%%%%%%%%%%%%%%%%%%%%%%%%%%%%%%%
\begin{figure}
    \centering
    \includegraphics[width=\linewidth]{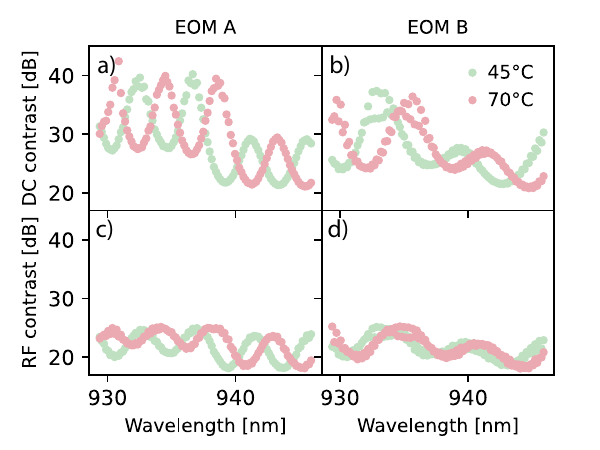} 
    \caption{Measured DC (a, b) and RF (c,d) contrast for the two different EOMs, as a function of wavelength, for two temperatures. The DC contrast is determined by a bias voltage sweep, and the RF pulse contrast is derived from the analysis of HBT measurements (see main text). A 10 dB fluctuation in contrast is visible over a wavelength change of a few nanometers. We see that these fluctuations can be shifted to longer wavelengths by heating up the device. The DC contrast is slightly higher than the RF contrast due to limited $V_{RF}$. For EOM B there is a fine structure around 935 nm, further measurements (not shown) uncover high frequency oscillations that are just resolved using the sample spacing in this figure.} 
    \label{fig:temperature} % MP note: eom A is Exail and eom B is iXblue
\end{figure}

With such HBT measurements we now characterize in detail the two different EOM devices A and B. In Fig. \ref{fig:temperature}, we show the DC and RF contrast of the EOMs as a function of laser (New Focus Velocity 3600 LN) wavelength, for different EOM temperatures (pink at $30{~^{\circ}}$C, green at $70{~^{\circ}}$C), which are temperature controlled using an attached restive heater. 

Panels (a) and (b) in Fig. \ref{fig:temperature} show that we obtain an excellent DC contrast above 20 dB ($I_p>100I_0T_0$), determined by the minimum and maximum transmission of a bias voltage sweep. The oscillatory features also appear in the RF contrast (panels (c) and (d)), but only up to 20 dB, indicating that the RF voltage does not fully reach $V_\pi.$ Care was taken that the same transmission minimum (Fig. \ref{fig:EOM-transmission}) was used for DC and RF contrast measurements.
We clearly see a strong wavelength dependence of the contrast, which fluctuates by $10$ dB over a wavelength range as little as $5$ nm. We also see that by changing the temperature it is possible to reach a high contrast at any wavelength within the operating range of the EOMs. 

As mentioned before, the splitting ratios of the directional couplers are crucial for the quality of the transmission minimum. When heating up the device, the geometry of the MZI and couplers changes, which impacts the MZI balance \cite{zook_temperature_1967}. To operate in the transmission minimum, it is thus crucial to set the temperature of the EOM to the optimum for the desired wavelength. We find that this method produces high-contrast pulses with long-term stability. This again emphasizes that matching the RF peak voltage to $V_\pi$ is not the most important aspect in making high contrast pulses using an EOM, but the minimum transmission strongly determines the achievable contrast (Fig. \ref{fig:EOM-transmission}).

%%%%%%%%%%%%%%%%%%%%%%%%%%%%%%%%%%%%%%%%%%%%%%%%%%%%%%%%%

\section{EOM based optical cross-correlation}\label{sec:xcor}
\begin{figure}
    \centering
    \includegraphics[width=\linewidth]{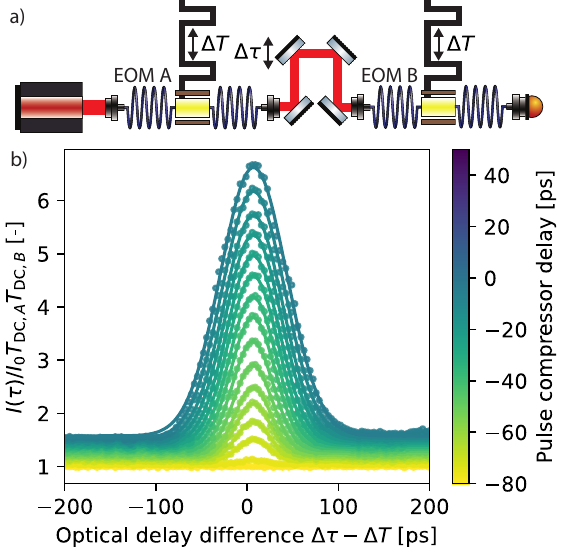} 
    \caption{Panel a) shows a schematic setup of the cross-correlation gating technique. A laser is sent to EOM A, then via an optical delay line to EOM B, the result is measured using a slow photo diode. Both EOMs are synchronously driven with a repetition rate that matches to the optical delay.
    Panel b) shows a measured cross-correlation when driving the EOMs without an amplifier, the through line are Gaussian fits. 
    The color of the line corresponds to the relative delay set on the pulse compressor, ranging from $-80$ (yellow) ps to $+50$ (blue) ps. As the delay difference increases, the pulses become longer as evidenced by the width of the peak.}
    \label{fig:autocorrelated_pulse}
\end{figure}

In HBT measurements, limited timing resolution and detector jitter does not allow to precisely determine the optical pulse shape. Here we introduce a cross-correlation method using two similar EOM devices operating in series via an optical delay line, shown schematically in Fig. \ref{fig:autocorrelated_pulse} (a). Importantly, this method only needs a slow photodiode. The optical power transmitted through the EOMs is given by the time averaged product of the transmission through EOM A(B): $T_{A(B)}(t).$ When the optical delay $\Delta\tau$ is adjusted to (a multiple of) the period of the pulse generator $\Delta T$, the correlation between the transmission from the EOMs can be measured by varying the optical delay as
\begin{equation}
    I(\tau) = \frac{1}{\Delta T} \int_{\Delta T} I_0 T_A(t) T_B(t-\Delta\tau) \mathrm{d}t.\label{eq:eom_autocor}
\end{equation} 
This method only requires moderately similar EOM characteristics and an electronic jitter lower than the pulse width.

The transmission of the EOM can be split into a part describing the DC bias $T_{\text{DC}}$ and the RF modulation signal $T_{\text{RF}}.$ If the DC bias of an EOM is chosen such that $T_\text{DC}$ is in a minimum, any non-zero RF modulation will result in an increased transmission. 
The normalized EOM transmission is obtained by dividing by the measured intensity without RF modulation
$C_{A(B)}(t)=T_{\text{RF},{A(B)}}(t)/T_{\text{DC},{A(B)}}$:
\begin{equation}
\begin{split}
    I(\tau)/\langle{I_0T_{\text{DC},A}T_{\text{DC},B}}\rangle =  \\1+ \langle{C_A(t)}\rangle+ \langle{C_B(t)}\rangle\\
    + \frac{1}{\Delta T}\int_{\Delta T} C_{A}(t)C_{B}(t-\Delta\tau) \mathrm{d}t.\label{eq:eom_autocor_DCRF}
\end{split}
\end{equation}
For the experimental demonstration of the correlation method, we first connect the EOMs directly to the pulse compressor without amplifiers. Fig. \ref{fig:autocorrelated_pulse} (b) shows the result for several settings of the relative delay of the electronics delay lines that determine the pulse length. We observe a Gaussian-shaped envelope of the cross-correlation signal, which is due to the integral term in Eq. \ref{eq:eom_autocor_DCRF}. The vertical offset between the curves is caused by the increase in time-averaged transmission with longer pulse lengths, which leaks through the other EOM due to finite $T_0$ - this is described by $\langle{C_A(t)}\rangle+ \langle{C_B(t)}\rangle$ in Eq. \ref{eq:eom_autocor_DCRF}.

The data in Fig. \ref{fig:autocorrelated_pulse} (b) is interpreted as the cross-correlation of the EOM transmission and is fit with a Gaussian function with a constant offset.
From this, the pulse height and width produced by a single EOM can be found. 
For a relative pulse compressor delay of $-75$ ps, corresponding to the condition of the spectrum analyzer measurement in Fig. \ref{fig:spectrum-analyzer}, the optical pulse length in the cross-correlation measurement is $24.2\pm1.8$ ps; this agrees very well to the spectrum analyzer result of $23.1\pm0.6$ ps.
We note, as the pulse length becomes long relative to the edge (pulse compressor delay of $\geq -20$ ps in Fig. \ref{fig:autocorrelated_pulse} (b)) the cross-correlation shape changes to a more triangular shape, as expected for square pulses, but is still well approximated by a Gaussian. 

At the maxima of the curves in Fig. \ref{fig:autocorrelated_pulse} (b) the two EOMs are exactly synchronized. In this case, the time dependent transmission of each EOM is multiplied: for Gaussian pulses, the transmission through the cascaded EOMs is $e^{-t^2/\sigma^2}e^{-t^2/\sigma^2}=e^{-2t^2/\sigma^2}.$ The pulse width (proportional to $\sigma$) is a factor $\sqrt{2}$ shorter than using a single EOM \cite{bhattacharya_picosecond_2014}, allowing an optical pulse down to $17$ ps. Additionally, the contrast of both EOMs is multiplied.

%%%%%%%%%%%

\begin{figure}
    \centering
    \includegraphics[width=\linewidth]{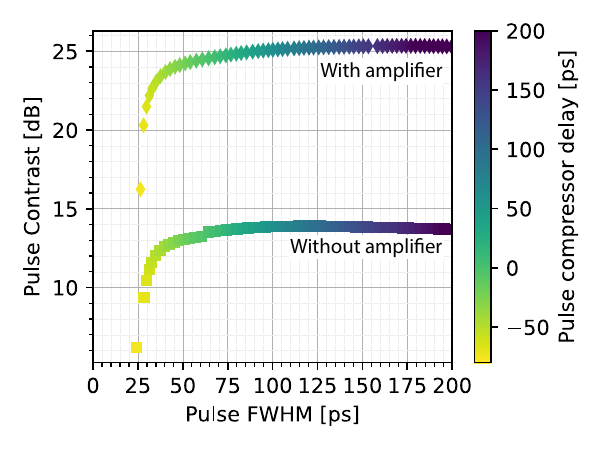}
    \caption{Pulse width (horizontal axis) and contrast (vertical axis) for different settings of the relative delay of the pulse compressor (color scale) without (square symbols) and with (diamond symbols) an amplifier between the pulse compressor output and the EOMs. Error bars smaller than the symbol size are not shown.}
    \label{fig:pulse_length_height}
\end{figure}
We now explore the trade-off between pulse contrast and pulse length with the cross-correlation technique. In Fig. \ref{fig:pulse_length_height} we compare pulses made with and without amplifier for several pulse compressor delay settings (color scale). The pulse contrast is rather low for short pulses of around $25$ ps, but rapidly increases and starts to saturate at $35$ ps long pulses. This inflection point is around a pulse compressor relative delay of $-50$ ps. Without amplifier the pulse contrast at saturation is $11$ dB, while with amplifier it is $22$ dB. Using an amplifier increases the pulse contrast for all pulse lengths by $~11$ dB without significantly affecting the pulse width, indicating that the system is not limited by the amplifier bandwidth.

%%%%%%%%%%%%%%%%%%%%%%%%%%%%%%%%%%%%%%%%%%%%%%%%%%%%%%

\section{EOM-driven single photon source}
Now we demonstrate the use of our EOM laser pulses to drive an InGaAs QD-microcavity based single photon source, see Ref. \cite{steindl_artificial_2021} for details about the device design. We use resonant excitation of a negatively-charged QD (lifetime $1/\gamma=710\pm110$ ps), and now different single photon detectors with a higher detection efficiency but 350 ps jitter. We align the device and optimize cross-polarization \cite{steindl_cross-polarization-extinction_2023} to remove the excitation laser using continuous-wave laser light, before we enable the EOMs to obtain short pulses. 

\begin{figure}
    \centering
    \includegraphics[width=\linewidth]{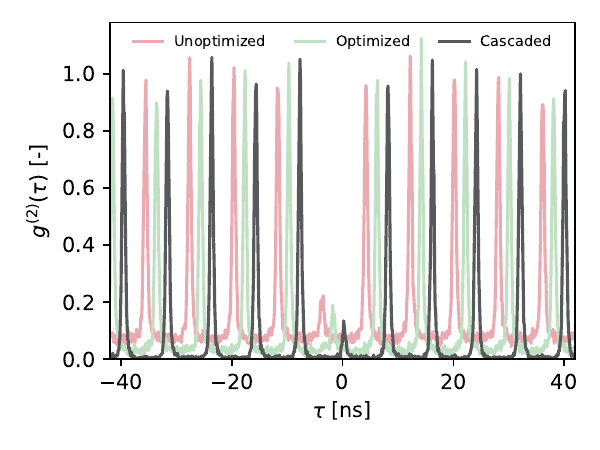}
    \caption{Measured $g^{(2)}(\tau)$ with a temperature unoptimized EOM ($40 {~^{\circ}}$C), a temperature optimized EOM ($27{~^{\circ}}$C) and a double optimized EOM (cascaded). We added a time (horizontal, $\tau$) offset between the measurements for clarity, the $g^{(2)}(\tau)$ axis is without added offset.}
    \label{fig:singlephotons}
\end{figure}

First, we measure the second-order correlation function $g^{(2)}(\tau)$ for various EOM pulser conditions, shown in Fig. \ref{fig:singlephotons}. For these measurements we use the same pulse length of $60\pm1$ ps, because this is the shortest pulse length that gives a good contrast in $g^{(2)}(\tau)$ for all conditions. We see that by EOM temperature optimization $g^{(2)}(0)$ decreases from $0.34\pm0.01$ to $0.27\pm0.01,$ and even more by cascaded operation of two temperature-optimized EOMs to $g^{(2)}(0)=0.18\pm0.01.$ This is as expected, since a reduced excitation laser background $I_0 T_0$ reduces the influence of QD re-excitation and detector jitter. This laser background makes it difficult to compare the optical excitation power, for the measurements shown in Fig. \ref{fig:singlephotons} the time-averaged power was kept constant; the pulse energy was kept below the energy required for full population inversion. In the cascaded case, passage through the second EOM further shortens the pulse length as mentioned above, which further decreases $g^{(2)}(0).$

We note that at a pulse length of $\sim0.1\gamma$ we should expect $g^{(2)}(0)\sim10^{-2}$ \cite{hanschke_quantum_2018}, which is similar to the value expected from our cross-polarization laser leakage \cite{steindl_cross-polarization-extinction_2023}. The higher observed value of $g^{(2)}(0)$ is most likely due to cavity-enhanced non-resonant emission \cite{suffczynskiOriginOpticalEmission2009, hohenesterPhononassistedTransitionsQuantum2009}.

Similarly as we have determined in Section \ref{sec:eomandcompressor} the laser pulse contrast from HBT measurements, we now determine the {\em single-photon} pulse contrast. For this, we follow the procedure explained above and in Appendix \ref{appx:hbt_analysis}, but ignoring data around $\tau=0$ because this peak is of course absent for single-photon sources. For a single EOM and without temperature optimization, we obtain (Fig. \ref{fig:singlephotons}) a single-photon pulse contrast of $22.8\pm0.1$ dB. This improves to $28.8\pm0.1$ dB by EOM temperature optimization, and if two temperature-tuned EOMs are cascaded, we observe a single-photon pulse contrast of $35.6\pm0.1$ dB.

\begin{figure}
    \centering
    \includegraphics[width=\linewidth]{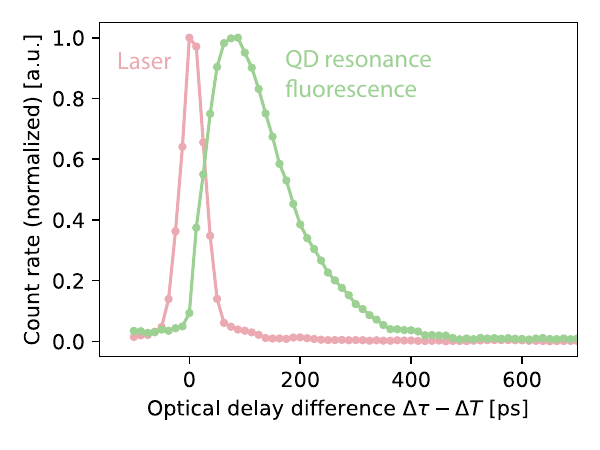}
    \caption{EOM-gated measurement of the QD resonance fluorescence and laser excitation pulse. We normalized the counts for ease of comparison.}
    \label{fig:xcor_qdot}
\end{figure}

Finally, we modify the cross-correlation setup of Sec. \ref{sec:xcor} to characterize the time-dependent resonance fluorescence of a (different) quantum dot: the first EOM is used to create short light pulses which excite the QD trion transition, and the second EOM is used to temporally gate the reflected light from the QD device, detecting integrated counts using a single photon detector with $350$ ps jitter. Now we use a relative pulse compressor delay of $-60$ ps, corresponding to an optical pulse length of $31.39\pm0.04$ ps.

We first measure the excitation laser (red curve in Fig. \ref{fig:xcor_qdot}) by tuning the quantum dot out of resonance and tuning the polarization condition. From this we obtain that the measured length of the transmission window of an EOM is $35.8\pm0.7$ ps. This window is $4$ ps longer than the previously measured single-EOM pulse length in Fig. \ref{fig:pulse_length_height}. We think that this small deviation is due to FPGA trigger signal jitter; this is more important here since the optical path between the EOMs is much longer ($\sim 20$ m extra fiber to the QD setup and back).

Now we measure QD resonance fluorescence, we see that the green curve in Figure \ref{fig:xcor_qdot} sharply increases to a maximum that coincides with the end of the excitation pulse. After the excitation pulse, the resonance fluorescence shows an exponential decay with a lifetime of $89\pm2$ ps. This allows us to resolve the cavity-enhanced quantum dot lifetime even though the end-to-end jitter of our single photon detection system is a few hundred picoseconds.

\section{Conclusion}
We have shown that a custom-made ps-range electronic pulse compressor in combination with integrated-optics EOMs can be used to drive a fast quantum-dot cavity single photon source.
By driving two EOMs with the same pulse compressor, combined with an optical delay, the pulse shape or the QD resonance fluorescence can be investigated with picosecond resolution with a slow detection system. Further, by cascading and synchronizing the EOMs the pulse length can be decreased down to $17$ ps while the single-photon pulse contrast increases by $\sim10$ dB. 
For achieving high contrast pulses it is most important to operate the EOM as close to the transmission minimum as possible, more so than reaching full modulation depth. The minimum transmission depends sensitively on the used wavelength and device temperature, both parameters can be optimized. In the future, new DC bias locking techniques \cite{zhang_fpga-based_2023, wang_versatile_2010} could further improve the pulse contrast, and we envision the generation of more complex pulse patterns.

\begin{acknowledgments}
We acknowledge funding from NWO/OCW (Quantum Software Consortium, No. 024.003.037), from the Dutch Ministry of Economic Affairs (Quantum Delta NL), and from the European Union’s Horizon 2020 research and innovation program under Grant Agreement No. 862035 (QLUSTER)

\end{acknowledgments}

%%%%%%%%%%%%%%%%%%%%%%%%%%%%%%%%%%%%%%%%%%%%%%%%%%%%%%%%%%%%%%%%%%%%%%%%%%%%%%%%

\bibliography{total_bib-20240326}% Produces the bibliography via BibTeX.
\bibliographystyle{naturemagwV1allauthors}

\appendix

\section{Pulse contrast from HBT coincidence counts}\label{appx:hbt_analysis}

The expected measured intensity $I(t)$ of the detector can be described as
\begin{equation}
    I(t)=I_\mathrm{BG}+f(t),\label{eq:received_photons}
\end{equation}
where $I_\mathrm{BG}$ is the background intensity and the function $f(t)$ describes the pulse shape as measured by the detector, including the detector response. The number of measured photons in a pulse is independent of detection jitter: $\int f(t) dt= \langle N_p\rangle.$ 

Substituting the measured intensity (Eq. \ref{eq:received_photons}) in the coincidence function 
$\langle I(t) I(t-\tau)\rangle$ (where the brackets average over one period $\Delta T$) produces three terms:
\begin{align}
    I_\mathrm{BG}^2 T + 2I_\mathrm{BG} \langle N_p \rangle + \int_{\Delta T} f(t)f(t-\tau) \mathrm{d}t.
\end{align}
The first two terms are not dependent on $\tau$ and together are equal to the background in correlations, the integral term describes the correlation peak. 
Integrating the last term over one period in both $\tau$ and $t$ results in 
\begin{align}
    \int_{\Delta T} \int_{\Delta T} f(t)f(t-\tau) \mathrm{d}t \mathrm{d}\tau = \langle N_p \rangle^2,
    % \int_{\Delta T} \left[ \langle I(t) I(t-\tau)\rangle - I_b^2 T - 2 I_b N_p\right] d\tau = \langle N_p \rangle^2,
\end{align}
allowing the mean detected photons in a pulse to be found. 

The ratio $N_p/N_\mathrm{BG}$ is proportional to the transmission contrast via the duty cycle $D$ of the pulses $N_p/N_b \approx D T_p/T_b.$ The exact relation is dependent on the pulse shape; for a square pulse it is exact while a Gaussian pulse introduces a factor of $\frac{1}{2}\sqrt{\pi/\ln{2}},$ when the pulse width is measured by the FWHM.

From a measured $g^{(2)}(\tau)$ histogram as shown in the right panel of Fig. \ref{fig:pulses}(b) the contrast can be calculated. By fitting a Gaussian to the main peak, the background level can be determined with high resolution, which is useful for high contrasts when the average background coincidences are very low.

\end{document}